# Nonlinearity in nanomechanical cantilevers

L. G. Villanueva[1,2], R. B. Karabalin[1], M. H. Matheny[1], D. Chi[1], J. E. Sader[1,3], M. L. Roukes[1]

[1]*Kavli Nanoscience Institute and Departments of Physics, Applied Physics, and Bioengineering, California Institute of Technology, Pasadena, California 91125*

[2]*Department of Micro- and Nanotechnology, Technical University of Denmark, DK-2800 Kongens Lyngby, Denmark*

[3]*Department of Mathematics and Statistics, The University of Melbourne, Victoria 3010, Australia*

## ABSTRACT

Euler-Bernoulli beam theory is widely used to successfully predict the linear dynamics of micro- and nano-cantilever beams. However, its capacity to characterize the nonlinear dynamics of these devices has not yet been rigorously assessed, despite its use in nanoelectromechanical systems development. In this article, we report the first highly controlled measurements of the nonlinear response of nanomechanical cantilevers using an ultra-linear detection system. This is performed for an extensive range of devices to probe the validity of Euler-Bernoulli theory in the nonlinear regime. We find that its predictions deviate strongly from our measurements for the nonlinearity of the fundamental flexural mode, which show a systematic dependence on aspect ratio (length/width) together with random scatter. This contrasts with the second mode, which is always found to be in good agreement with theory. These findings underscore the delicate balance between inertial and geometric nonlinear effects in the fundamental mode, and strongly motivate further work to develop theories beyond the Euler-Bernoulli approximation.

# I. Introduction

Micro- and nanoelectromechanical systems (MEMS and NEMS) are increasingly spawning a wide range of sensing applications, including detection of mass[1,2], force[3], and spin[4]. In addition, they can also be used as time reference devices[5] and as basic tools to explore fundamental physical processes[6] and dynamical effects[7]. At small vibrational amplitudes these systems behave as linear mechanical devices. However as the amplitude increases, nonlinear effects are readily manifested[8,9]. This becomes of central importance in all of the aforementioned fields of application. For example, nonlinear phenomena impose a fundamental limit for the minimum detectable frequency shift[10], while simultaneously enabling rich and complex dynamical behavior[11].

Arguably the most utilized mathematical description of the deformation of MEMS and NEMS cantilever beams is Euler-Bernoulli theory[12]. We observe that this theory accurately ($\pm 2.5\%$) predicts the resonant frequencies and other linear parameters for the flexural vibration modes of thin cantilever beams of aspect ratios (AR = length/width) greater than two; see Appendix I. The generic Euler-Bernoulli theory implicitly assumes the beam to be one-dimensional and is formally valid in the asymptotic limit of infinite AR. For beams of finite AR and non-negligible thickness, it is sometimes necessary to include the effects of transverse[13] or shear[12] deformation respectively, although these effects are second-order and can be often ignored in experimental design and application[14-17].

As introduced before, nonlinear behavior manifests for finite amplitude of motion. This is true not only at the micro- and nano-scale, but also for macroscopic structures such as airplane wings[18,19]. Consequently, an effort to predict the dynamics of the nonlinear response and the parameters governing it has recently gained momentum[20-22].

Nonlinearity in the dynamic response of mechanical structures can have a multitude of origins[8,23], including transduction effects (actuation/detection)[21], material properties (nonlinear constitutive relations)[24], non-ideal boundary conditions[25,26], damping mechanisms[27,28], adsorption/desorption processes[29], and geometric/inertial effects[30,31]. Geometric nonlinearities can appear in any mechanical structure when large deformations induce a nonlinear relation between strain and curvature, thus modifying the effective stiffness of the structure. Inertial nonlinearities are typically induced through the generation of additional degrees of freedom in the motion, which serve to enhance the effective mass of the structure.

The intrinsic (i.e. originating from the mechanical structure) nonlinear response of doubly-clamped beams has been shown to be dominated by a geometric nonlinearity due to enhanced tension along the beam. Stiffening behavior is observed[5,32], which is accurately predicted by Euler-Bernoulli theory[8]. In contrast, the nonlinear response of cantilever beams has received comparatively little attention. Most articles report theoretical investigations of the nonlinear response of these structures[21,31,33]. These studies predict a stiffening nonlinearity for the fundamental mode, while the higher order modes are predicted to be softening in nature. Strikingly, experimental assessment of the validity of such calculations for the fundamental mode has been limited in geometrical range and statistical analysis, and has not provided measurements with linear transduction[20,21].

In this article, we address this issue and present detailed experimental results for the intrinsic nonlinear resonant responses of nanomechanical cantilever beams. In particular, we study the first and second flexural out-of-plane modes. The fabrication of these devices and the transduction of their motion are optimized in order to minimize the effect of other sources of nonlinearity. We show that Euler-Bernoulli beam theory yields predictions for the first mode that significantly deviate from our experimentally observed data, especially for cantilevers of low AR. In contrast, excellent agreement between theory and measurement is observed for the second mode. These results have significant implications for experimental design and interpretation, and are expected to stimulate further improvement in theoretical modeling beyond Euler-Bernoulli beam theory, as we discuss below.

## II. Theory

The type of structure that we use in our study is shown in the schematic of Fig. 1. Cantilever beams with U-shaped geometry are chosen given their interest for various applications[34,35] and to facilitate linear detection, as will be detailed later. The cantilever beams have a total length $L$ and width $b$. The region close to the clamp presents two legs of length $L_{leg}$ and width $b_{leg}$. In our particular examples, the structures are designed to have: $L = 3 \cdot L_{leg}$ and $b = 3 \cdot b_{leg}$. The linear dynamic analysis of these cantilevers, according to Euler-Bernoulli theory, is easily obtained using an analysis for beams with non-uniform cross sections[12,36]; see Appendix I.

The geometric and inertial nonlinearities in our cantilever structures according to Euler-Bernoulli theory are calculated using Hamilton's principle, the Galerkin method, and assume that only one normal mode is active[31,33]. This yields Eq. (1) for the dynamics of the $n^{th}$ mode, where we have omitted the index $n$ for simplicity:

$$m_{\text{eff}}\ddot{x} + \frac{m_{\text{eff}}\omega_R}{Q}\dot{x} + k_{\text{eff}}x + \frac{\beta_{\text{geom}}}{L^2}x^3 + \frac{\beta_{\text{iner}}}{L^2}(x\dot{x}^2 + x^2\ddot{x}) = G\cos(\omega t), \qquad (1)$$

where the dot denotes the time derivative, $Q$ is the quality factor, $G\cos(\omega t)$ is the externally applied driving force, and $m_{\text{eff}}$, $k_{\text{eff}}$, $\beta_{\text{geom}}$, $\beta_{\text{iner}}$ are the effective mass, effective elastic constant, geometrical nonlinear coefficient, and inertial nonlinear coefficient, respectively, and they are given by:

$$m_{\text{eff}} = \int_0^1 \mu(\xi)\phi(\xi)^2 \, d\xi, \qquad k_{\text{eff}} = \int_0^1 \langle EI\rangle(\xi)\phi''(\xi)^2 \, d\xi = m_{\text{eff}}\omega_R^2,$$

$$\beta_{\text{geom}} = 2\int_0^1 \langle EI\rangle(\xi)\big(\phi'(\xi)\phi''(\xi)\big)^2 \, d\xi, \qquad \beta_{\text{iner}} = \int_0^1 \mu(\xi)\left(\int_0^\xi \phi'(\zeta)^2 \, d\zeta\right)^2 d\xi, \tag{2}$$

where $\mu(\xi)$ is the mass per unit length as a function of normalized distance $\xi$ along the beam axis, $\langle EI\rangle(\xi)$ the bending rigidity, $\phi(\xi)$ is the normalized mode shape, and the primes denote spatial derivatives. Note that this theoretical formulation is generally applicable to cantilevers with spatially varying cross sections like the devices used in this work.

Using secular perturbation theory[8], we can solve Eq. (1) and extract the amplitude response in the vicinity of the resonant frequency $\omega_R$:

$$x^2(\omega) \approx \frac{\left(\frac{G}{2k_{\text{eff}}}\right)^2}{\left(\frac{\omega-\omega_R}{\omega_R} - \frac{3}{8}\frac{\alpha}{L^2}x^2(\omega)\right)^2 + \left(\frac{1}{2Q}\right)^2}, \tag{3}$$

where $\alpha$ is the dimensionless nonlinear coefficient, that depends on both inertial and geometric nonlinearity,

$$\alpha = \frac{\beta_{\text{geom}}}{k_{\text{eff}}} - \frac{2}{3}\frac{\beta_{\text{iner}}}{m_{\text{eff}}}. \tag{4}$$

Note that $\beta_{\text{geom}}, \beta_{\text{iner}} > 0$, and therefore the final nonlinearity of the structure is determined by two competing effects: geometric and inertial nonlinearities. The former stiffens the structure at large amplitudes while the latter leads to a softening effect. For the cantilevers used in this work, using the mode shapes that are derived in the Appendix I, we obtain: $\alpha_1 = 0.044 \pm 0.001$ and $\alpha_2 = -18.6 \pm 0.15$ for the first and second flexural modes, respectively. Variations in the parameters $\alpha_1$ and $\alpha_2$ are due to fabrication non-uniformities, as will be discussed below.

To experimentally assess the validity of these calculations, we utilize a system that employs a highly linear transduction technique to actuate and detect the motion. The resonators are made from well-characterized materials, allowing us to stay within their linear range of mechanical response. Also, the fabrication process (based on bulk-micromachining) yields cantilever beams with well-defined clamping regions.

## III. Fabrication

The fabrication of the devices starts with 725 μm double sided polished, 100 mm in diameter, silicon wafers. We deposit a 500 nm thick layer of low stress LPCVD (low pressure chemical vapor deposition) silicon nitride (SiN) on both sides of the wafer; Fig. 2(a). We then pattern the SiN on one side of the wafer (backside) using photolithography and dry-etching, prior to performing an anisotropic silicon etching in KOH (potassium hydroxide); Fig. 2(b). This step defines SiN membranes on one side of the wafer (front side).

Once the membranes are defined, we perform electron beam lithography using a double layer of PMMA in order to lift-off the metal layer. We evaporate a bilayer of Cr (5 nm – adhesion layer) and Au (50 nm) which is subsequently patterned using the lift-off of the PMMA double layer processed before; Fig. 2(c). A second lithography and lift-off process is then performed to define the metal contacts with a thicker metal layer (Cr/Au, 5/150nm).

Finally, using the gold as a hard mask, we perform a mild dry etching of the silicon nitride layer, which defines the released structures with no undercut at the clamping region (Fig. 2.d, Fig. 1). The resulting structures are a tri-layer stack of SiN ($510 \pm 5$ nm thick), chromium (adhesion layer, 5 nm), and gold ($25 \pm 5$ nm). Note that the final gold layer has a decreased thickness as a consequence of the dry etching that is performed.

Some examples of the released structures are shown in Fig. 3 (a) and (b), where we can see the two legs near the clamp that permit sensitive detection of the cantilever motion. The structures are designed to have a width of $b = 4.5$ µm, and the width of each leg to be $b_{leg} = 1.5$ µm. Deviations from these values between devices are of order ±50 nm. A range of cantilevers of different lengths is fabricated, with AR ranging from 2 to 13, and the legs designed to be one third of the total length. Alignment tolerance causes dispersion of approximately ±1 µm in the total cantilever length and the length of the legs (an example of this can be seen in Appendix III). This variation affects slightly the theoretical estimation of the nonlinear coefficients. Therefore, every device is individually inspected using a Scanning Electron Microscope (SEM) to accurately determine the dimensions and hence the theoretical nonlinear coefficients.

## IV. Experimental results

Actuation is performed by means of a piezoshaker stage operating in linear regime, taking precautions against the effects of electric leakage. Measurements are performed using two techniques: (i) a highly sensitive optical detection scheme for the detection of very small amplitudes (e.g. thermomechanical noise); (ii) the highly linear metal-based piezoresistive readout technique[32,37] for the detection of larger amplitudes.

Thermomechanical noise data is used to calibrate the optical detection responsivity ($V_{optical}$/m) in its linear range, for small amplitudes, using the equipartition theorem. We then obtain the metal-based piezoresistive detection responsivity ($V_{PZM}$/m) by comparing the resonant responses for both readout methods, keeping the drive levels low to maintain linearity of the optical detection. Finally, large amplitude motion and nonlinear behavior are captured using metal-based piezoresistive detection[37], which is linear and now calibrated. More details on the experimental procedure are shown in the Appendix II. We would like to emphasize here that calibration of the motion is not performed using the nonlinear coefficient, as has been proposed in the past[38], but using an independent phenomenon: the Brownian motion of the cantilevers. Examples of the observed dynamic responses for different drives are shown in Fig. 3(c)-(f), for the first and second flexural modes and for two cantilever devices of different AR. While the second mode presents a softening nonlinearity in both cases Fig. 3(e)-(f), the behavior of the first mode can vary from being a stiffening nonlinearity [Fig. 3(d)] to one that is softening [Fig. 3(c)], as the AR is varied.

To facilitate quantitative comparison between different devices, we extract the dimensionless parameter $\alpha$ from the measurements using a double fitting procedure: (i) we first fit the full-resonant response to Eq. (3), and then (ii) we fit the frequency positions of the maxima, $\omega_{max}$, for each drive to $(\omega_{max} - \omega_R)/\omega_R - 3\alpha x_{max}^2/(8L^2)$. These two procedures yield the same parameter values, and thus provide a consistency check on fit procedure robustness. The values for $\alpha$ for the first mode are given in Fig. 4(a) whereas those for the second mode are in Fig. 4(b). Both figures display a solid grey line that denotes the predicted theoretical value from Euler-Bernoulli beam theory, taking into account the non-constant cross section; see Section II. For the two smallest ARs (i.e. AR = 2 and 3) the thermomechanical motion of the second mode cannot be detected given the high stiffness of those modes, and thus their nonlinearity cannot be characterized. The same was true for modes higher than the second. Nonetheless, such higher order modes have been measured previously on macroscale devices[23,39], yielding good agreement with Euler-Bernoulli theory.

Figure 4 displays the differences in the nonlinear behavior exhibited by the first two cantilever modes. We summarize the observations:

*First mode*: Figure 4(a) clearly shows a systematic decrease in the nonlinear parameter $\alpha$ for the fundamental mode, as AR is reduced, with the behavior changing from stiffening to softening – experimental values approach the theoretical (stiffening) prediction for large AR. The solid line gives the theoretical prediction of Euler-Bernoulli theory; the dotted line delineates softening and stiffening behavior; and the dashed line is presented only as a visual aid.

*Second mode*: The experimental data in Fig. 4(b) contrast strongly with those for the first mode [Fig. 4(a)]. The dashed line represents the average of the experimental data and the boundaries of the colored zone define one measured standard deviation from the mean. No dependence on AR is observed, with the theoretically calculated value deviating by less than 5% from the mean and within one standard deviation.

## V. Discussion

To highlight the qualitative and quantitative differences between the two modes of vibration investigated, results for the relative difference between the theoretically predicted and experimentally observed values are given in Fig. 4(c). This clearly demonstrates that significant deviations exist in the first mode at low AR, whereas excellent agreement is always achieved for the second mode. Experimental data for the first mode are in reasonable agreement with theory for AR greater than ten, while the relative difference exceeds a factor of 10 when the AR is two. This difference between our experimental results and the theoretical estimations (for non-uniform beams) provides clear evidence that a breakdown in the underlying assumptions of Euler-Bernoulli beam theory is behind the observed variation with respect to AR. It is important to note that the Euler-Bernoulli formula Eq. (2), which includes existence of the cantilever legs, predicts that the sign of the nonlinear coefficient does not change for the first mode, regardless of leg width and/or length, i.e. the predicted nonlinearity is independent of Aspect Ratio.

We note that material nonlinearity and fabrication uncertainties such as surface roughness, clamping variations, surface damage, fabrication residues etc., could lead to deviations in the device dimensions and material properties, which, in turn, could affect the overall nonlinear response. Some of these effects might be randomly distributed and may be responsible for the observed scatter in measurements [Fig. 4(b)]. However, these effects are not expected to lead to the observed systematic deviation in the nonlinear response for the first mode as a function of AR. Shear deformation effect is estimated to be negligible, due to the large length/thickness ratios ($L/h \sim 18 - 180$).

Nonlinearity in Euler-Bernoulli beam theory, as discussed above, emerges from two competing mechanisms: inertial and geometrical nonlinearities. The first mode exhibits individual geometric and inertial nonlinearities of nearly identical magnitude – their difference, and hence the overall nonlinearity, is an order of magnitude smaller than their individual contributions, according to Eq. (2). Thus, the presence of any additional and unspecified (small) nonlinearity can potentially modify the overall nonlinear response. This delicate balance is illustrated in Fig. 5, where we present measurements of the resonant behavior of two cantilevers with the same design dimensions (AR is 7, and their SEM images are shown in Appendix III). Strikingly, for the first mode, one device displays a stiffening response whereas the other device is softening [see Fig. 5(a) and (b)]. This anomalous behavior is in direct contrast to the second mode, which displays a definitive softening nonlinearity that is quantitatively identical for both devices [see Fig. 5(c) and (d)].

We now outline possible mechanisms driving the observed anomalous behavior for the first mode. For reference, we initially consider the second mode: Euler-Bernoulli beam theory predicts that the inertial nonlinearity significantly dominates the geometric term, leading to an overall softening nonlinearity. Our measurements yield good quantitative agreement with this theory. No dependence on AR is observed. This demonstrates that the inertial nonlinearity for the second mode is weakly dependent or insensitive to AR and well described by Euler-Bernoulli theory.

Reduction in AR or increase in mode number leads to a breakdown in a fundamental tenet of Euler-Bernoulli theory: a uniaxial stress distribution along the beam. Since we do not observe any AR dependence in the inertial nonlinearity of the second mode, we then conclude that the first mode inertial nonlinearity is also insensitive to AR.

For the first mode, when the reduction of AR causes a deviation from an uniaxial stress distribution, it may induce either (i) modification in the nonlinear stiffness term alone in the first mode, or (ii) both nonlinear stiffness and inertia being slightly affected, and thus tipping the fine balance between these terms. Either possibility can contribute to the observed enhanced softening with decreasing AR. Higher order cross-coupling between these terms may also be responsible. Importantly, the precise mechanism can only be discerned through use of theories beyond Euler-Bernoulli, which account for the complex stress distribution in higher dimensional elastic bodies[13,40,41].

## VI. Conclusions

In conclusion, careful fabrication and characterization enable us to experimentally measure the nonlinear dynamics of nanomechanical cantilevers as a function of their AR (length over width). This allows us to carry out the first detailed assessment of the validity of Euler-Bernoulli beam theory to describe the nonlinear response of these widely-used devices. Our study clearly demonstrates the validity of this theory for the second flexural mode of vibration, regardless of AR. However, this theory is incapable of properly describing our experimental data for the fundamental (first) flexural mode. Both softening and stiffening behaviors are observed for devices with identical geometries, and a systematic trend of enhanced softening evolves with decreasing AR. These findings strongly motivate development of theories beyond the Euler-Bernoulli approximation, which does not properly describe the nonlinear dynamics of the first flexural mode. They are also of fundamental importance in design and interpretation of nonlinear measurements that make use of nanomechanical cantilever devices.

## Acknowledgements

We would like to thank I. Bargatin and E. Myers for useful suggestions and discussions. L.G.V. acknowledges financial support from the European Commission (PIOF-GA-2008-220682) and Prof. A. Boisen. J.E.S. acknowledges support from the Australian Research Council grants scheme.

## Appendix I – Euler-Bernoulli for non-constant cross-sections

To determine the linear resonant frequency of cantilevers with non-uniform cross section, we use Euler-Bernoulli theory. For the cantilevers studied, see Fig. 1, there are two distinct zones of different but constant width. The governing equation for each zone is therefore

$$\langle EI \rangle_i \frac{d^4 Z(x,t)}{dx^4} + \mu_i \frac{d^2 Z(x,t)}{dt^2} = 0, \qquad (5)$$

where $Z(x,t)$ is the out-of-plane deflection as a function of the longitudinal coordinate within the beam, $x$, and time, $t$. The functions $\langle EI \rangle_i$ and $\mu_i$ are the (constant) flexural rigidity and (constant) mass per unit length, respectively, of each zone. The subscript $i$ indicates that the variable is zone dependent.

To proceed, $Z(x,t)$ is expressed in terms of the explicit time dependence, $\cos(\omega_n t)$, such that

$$Z(x,t) = C_n \phi_n(x) \cos(\omega_n t); \qquad (6)$$

where $i$ is the usual imaginary unit, $\phi_n(x)$ is the mode shape of mode $n$, and $\omega_n$ is the required linear frequency.

The required boundary conditions are:

$$\phi_n(x=0) = 0, \quad \phi'_n(x=0) = 0,$$

$$\phi''_n(x=L) = 0, \quad \phi'''_n(x=L) = 0,$$

$$\phi_n(x \to L^-_{leg}) = \phi_n(x \to L^+_{leg}), \quad \phi'_n(x \to L^-_{leg}) = \phi'_n(x \to L^+_{leg}),$$

$$2b_{leg}\phi''_n(x \to L^-_{leg}) = b\phi''_n(x \to L^+_{leg}), \quad 2b_{leg}\phi'''_n(x \to L^-_{leg}) = b\phi'''_n(x \to L^+_{leg}),$$

which ensure continuity of the mode shape, its slope, moment and force between the two zones.

This system presents an eigenvalue problem, which can be solved analytically given the tractability of Eq. (5). Nonetheless, the resulting solution is complicated given the number of boundary conditions and the requirement to solve two 4$^{th}$ order differential equations and match their solutions. This analytical solution was therefore obtained using Mathematica®.

Table A1 compares the predictions of Euler-Bernoulli theory to results from a full three-dimensional finite element analysis, for the cantilevers studied. Actual dimensions of each individual device are measured using a Scanning Electron Microscope after the experiments are performed. Material properties are $E_{SiN} = 250$ GPa, $v_{SiN} = 0.27$, $\rho_{SiN} = 3440 \frac{kg}{m^3}$ for silicon nitride; $E_{Au} = 78$ GPa, $v_{Au} = 0.44$, $\rho_{Au} = 19300 \frac{kg}{m^3}$ for gold; and $E_{Cr} = 280$ GPa, $v_{Cr} = 0.21$, $\rho_{Cr} = 7190 \frac{kg}{m^3}$ for chromium. Here, $E$ is Young's modulus, $v$ is Poisson's ratio, $\rho$ is density, where the subscript indicates the material. The utilized mesh is refined until 99.9% convergence in the resonance frequency is achieved. The gold layer thickness (nominally 50 nm) was adjusted to ensure agreement between measurement and finite element analysis, i.e., one single thickness (25 nm) was used for all the simulations. It is striking that Euler-Bernoulli theory accurately predicts the full 3D FEM simulation results to within 2.5%.

## Appendix II – Experimental protocol

The experimental protocol to determine the nonlinear coefficients is based on optical calibration of the metal-based piezoresistive (piezometallic/PZM) detection scheme, which is linear over a large range of amplitudes.

### 1. Optical detection calibration

The first step utilizes a highly sensitive optical detection method to measure the Brownian motion of the mechanical device, i.e. its thermomechanical noise. This is achieved using an optical interferometer with a laser focused on the device. Application of the equipartition theorem then enables the responsivity of the optical detection scheme, $\Re_{\text{optical}} = V_{\text{optical}}/\text{nm}$, to be calibrated *at low amplitudes*, i.e. the range where the optical detection is still linear: $V_{\text{optical}}$ is the voltage output from the optical interferometer. Note that this optical responsivity, $\Re_{\text{optical}}$, is only used to calibrate the deflection at low amplitudes, because optical interferometric detection becomes nonlinear at moderate to large amplitudes (see next section and Fig. 5).

### 2. Piezometallic detection calibration

Piezometallic detection is a highly linear method that has been used widely in accelerometers, pressure sensors and control instruments[42]. However, for the devices used in this article it is not possible to observe their Brownian motion directly with this detection technique. Fortunately, there is a range of amplitudes for which piezometallic detection can be used and optical interferometric detection remains linear. We utilize this favorable small amplitude range to determine the responsivity of the piezometallic detection scheme. This is achieved using the following relations:

$$\mathfrak{R}_{\text{PZM}} = \frac{V_{\text{PZM}}}{nm} = \frac{V_{\text{PZM}}}{V_{\text{optical}}} \frac{V_{\text{optical}}}{nm} = \frac{V_{\text{PZM}}}{V_{\text{optical}}} \mathfrak{R}_{\text{optical}}, \qquad (9)$$

where $V_{\text{PZM}}$ is the voltage output from the piezometallic (PZM) detection. We estimate the ratio $V_{\text{PZM}}/V_{\text{optical}}$ by driving each cantilever at a given amplitude and detecting its motion using both techniques.

### 3. Large amplitudes detection using piezometallic detection

The dynamic range (linear response) of the piezometallic detection scheme is very large, and therefore its responsivity $\mathfrak{R}_{\text{PZM}}$ remains constant over several orders of magnitude in displacement. This range encompasses all displacements measured in this study. This linear detection scheme thus enables all cantilever nonlinear effects to be captured accurately.

### 4. Actuation

Actuation is performed using a piezoshaker ceramic attached to the bottom of the silicon chips containing the cantilevers. Due to the high quality factor of the devices ($Q \sim 1000 - 3000$), high voltages are not needed to actuate the piezoshaker to achieve a nonlinear mechanical response in the cantilevers. The maximum voltage that was applied (1 $V_{\text{rms}}$) generated an electric field three orders of magnitude below the reported onset of piezoelectric nonlinearity for the piezoshaker used.

To avoid electrical leakage that might affect the nonlinear response via a gate effect, the top-plane of the piezoshaker is always grounded, and AC power is applied to the bottom-plane of the piezoshaker. Also, we physically position our devices so that the defined grounding plane shields them from any gate effect.

## Appendix III – Device dimensions

As described in Section III, the devices are fabricated using a combination of Electron Beam Lithography (EBL) on the front side of the wafer (to pattern the shape of the cantilevers) and optical lithography on the backside (to define the membranes where the cantilevers reside).

Due to this combination of optical and electron beam lithography an excellent alignment tolerance of only ±1 μm exists. This leads to dispersion in the total cantilever length, $L_{\text{total}}$, and the legs length, $L_{\text{legs}}$, of ±1 μm. The width of the cantilevers (both total and leg widths) is defined to within ±50 nm. The dimensions of each device are measured using SEM after characterizing their nonlinear response, and these dimensions are used in all theoretical calculations. Predictions for the resonant frequencies are within 5% of the experimental values, based on known material properties and dimensions. We believe that such dispersion is a direct consequence of various fabrication uncertainties present during our process, such as surface roughness, surface damage, polymer residues etc.

Figure 5 gives the nonlinear response of two devices that were designed to have the identical dimensions. Due to the above-described misalignment, cantilever lengths inevitably differ slightly. In addition, there might be some incommensurable differences due to surface roughness, surface damage, polymer residues etc. Fig. 6 shows scanning electron micrographs of the actual two devices used in Fig. 5, with an identical aspect ratio of 7; qualitatively different nonlinear response for these two devices was observed (a: hardening; b: softening).

# Figures

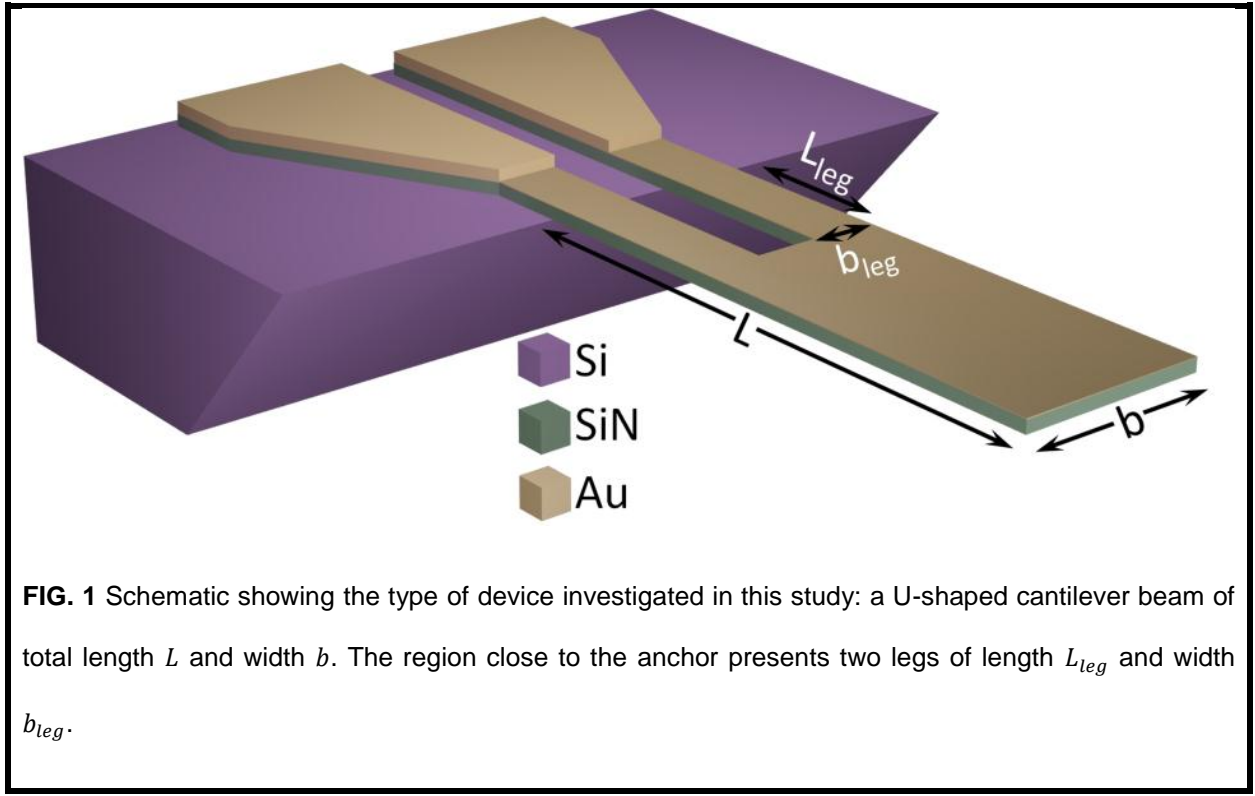

**FIG. 1** Schematic showing the type of device investigated in this study: a U-shaped cantilever beam of total length $L$ and width $b$. The region close to the anchor presents two legs of length $L_{leg}$ and width $b_{leg}$.

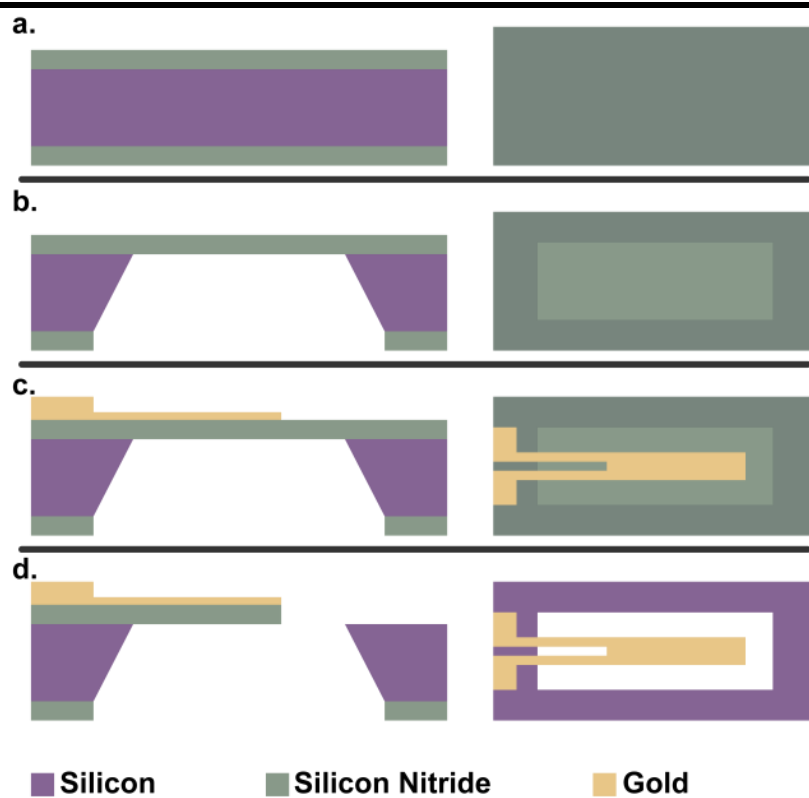

**FIG. 2** Schematic for the fabrication process flow. Side view is shown in the left column and corresponding top view is depicted in the right column. (a) SiN is deposited on both sides of a Si wafer. Backside SiN is patterned to define windows for the subsequent anisotropic silicon etching in KOH, yielding membranes on the front side (b). We then deposit (c) two bi-layers Cr/Au by means of two subsequent lift-off processes: one to be used in the detection of motion (5nm/50nm Cr/Au) and another one to define the contacts (5nm/150nm Cr/Au). (d) Using the gold as a hard mask, we perform a mild dry etching of the silicon nitride layer which defines the released structures with a proper clamping region, i.e. with no undercut (see also Fig. 1).

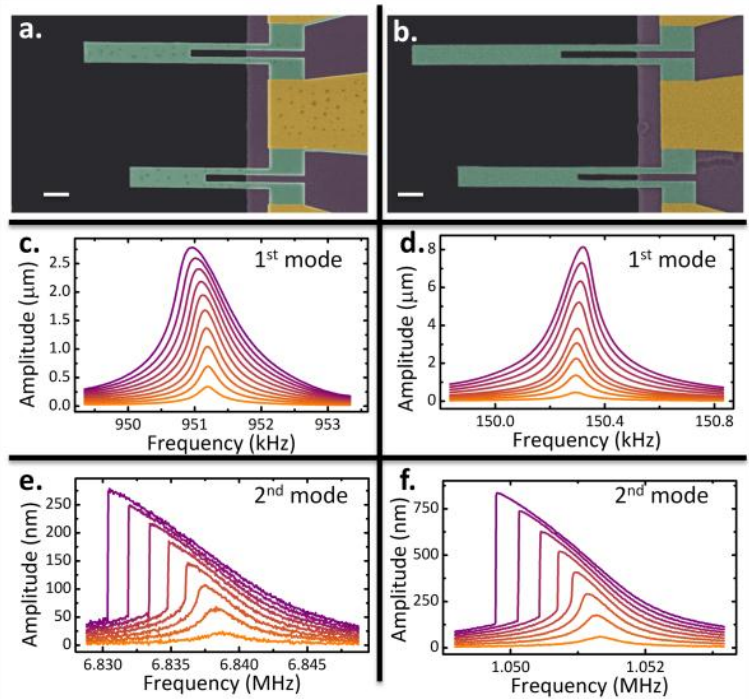

**FIG. 3** Examples of nanomechanical cantilever devices, (a) and (b), and their respective nonlinear responses for the first (c)-(d) and second (e)-(f) flexural vibrational modes. The micrographs show four structures with different ARs (AR, length over width): 5, 7 (a) and 12, 13 (b). The resonant responses (c)-(f) show the amplitude of vibration as a function of the drive frequency for different magnitudes of the driving force. The data correspond to cantilevers of AR 5 (c),(e) and AR 13 (d),(f). Scale bars are 5 µm.

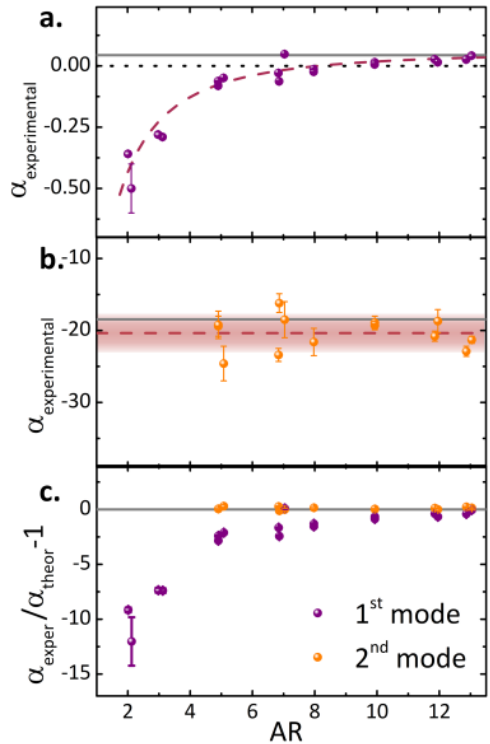

**FIG. 4** Experimentally estimated nonlinear coefficients for the first (a) and the second (b) out-of-plane vibrational modes. While data for the second mode (b) show good agreement with theory, data for the first mode (a) clearly diverge from the calculated value, mainly for low AR. This is highlighted when plotting the relative difference between experiment and theory (c). The first mode shows around one order of magnitude difference with the expected value for low AR. This difference is reduced for large AR. The second mode experimental measurements lie within some tens of percent of the expected value. A grey solid line represents the theoretical prediction in each plot. In (a) a dotted line delineates softening and stiffening behavior, while a dashed line is presented as a visual aid to follow the systematic experimental trend. In (b) the dashed line represents the average experimental value and the boundaries of the colored zone define one standard deviation from the mean.

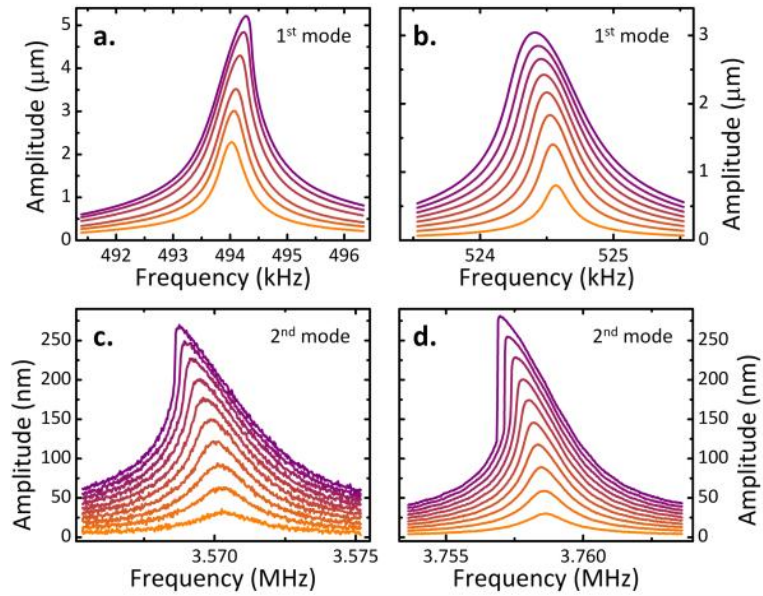

**FIG. 5** Resonant responses for different driving forces for the first (a, b) and second (c, d) mode of two cantilevers with the same AR in the design (AR 7 – due to alignment tolerances the real AR is 7.62, a-c, and 7.43, b-d). We show that the second mode presents softening nonlinearity on both cases, while the first mode presents both stiffening (a) and softening (b) nonlinearities. This may be due to additional unspecified effects such as material nonlinearity or fabrication-related differences (surface damage, polymeric residues, etc.).

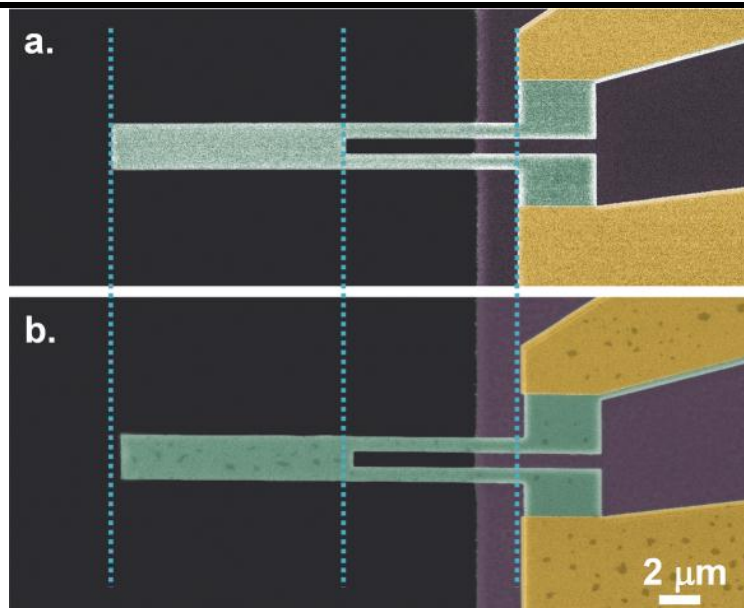

**FIG. 6** SEM micrographs showing two devices that were designed to have an identical aspect ratio of 7. Slight differences due to alignment mismatch can be observed. The nonlinear response of these device (a) is presented in Fig. 5(a) and (c); whilst the nonlinear response for device (b) is given in s is given in Fig. 5(b) and (d).

|  | 1st Mode | | | | 2nd Mode | | | |
| --- | --- | --- | --- | --- | --- | --- | --- | --- |
|  | $f_{exp}$ (MHz) | $f_{Sim}$ (MHz) | $f_{Th}$ (MHz) | $\left(\frac{f_{sim}}{f_{Th}}-1\right)100$ | $f_{exp}$ (MHz) | $f_{Sim}$ (MHz) | $f_{Th}$ (MHz) | $\left(\frac{f_{sim}}{f_{Th}}-1\right)100$ |
| 2 | 5.26 | 5.269 | 5.382 | 2.15 | | | | |
| 2 | 5.86 | 5.85 | 5.988 | 2.36 | | | | |
| 3 | 2.486 | 2.483 | 2.531 | 1.95 | | | | |
| 3 | 2.684 | 2.687 | 2.740 | 1.99 | | | | |
| 5 | 0.951 | 0.954 | 0.970 | 1.68 | 6.838 | 6.747 | 6.89 | 2.06 |
| 5 | 1.02 | 1.026 | 1.043 | 1.69 | 7.32 | 7.224 | 7.35 | 1.73 |
| 5 | 1.02 | 1.026 | 1.043 | 1.69 | 7.353 | 7.224 | 7.35 | 1.73 |
| 7 | 0.496 | 0.501 | 0.507 | 1.22 | 3.57 | 3.5435 | 3.60 | 1.48 |
| 7 | 0.525 | 0.528 | 0.534 | 1.19 | 3.76 | 3.72 | 3.76 | 1.17 |
| 7 | 0.527 | 0.531 | 0.538 | 1.38 | 3.8 | 3.75 | 3.79 | 1.00 |
| 8 | 0.39 | 0.392 | 0.396 | 0.98 | 2.815 | 2.77 | 2.79 | 0.77 |
| 8 | 0.387 | 0.392 | 0.396 | 0.98 | 2.803 | 2.77 | 2.79 | 0.77 |
| 10 | 0.25 | 0.253 | 0.255 | 0.95 | 1.81 | 1.79 | 1.80 | 0.62 |
| 10 | 0.25 | 0.253 | 0.255 | 0.95 | 1.807 | 1.79 | 1.80 | 0.62 |
| 12 | 0.178 | 0.1783 | 0.179 | 0.57 | 1.278 | 1.26 | 1.26 | 0.36 |
| 12 | 0.175 | 0.1752 | 0.176 | 0.69 | 1.238 | 1.24 | 1.24 | 0.32 |
| 13 | 0.15 | 0.1515 | 0.152 | 0.51 | 1.08 | 1.08 | 1.07 | 0.57 |
| 13 | 0.145 | 0.1473 | 0.148 | 0.53 | 1.051 | 1.05 | 1.05 | 0.43 |

**Table A1:** Comparison of resonant frequencies (MHz) for cantilevers studied. Results given for measurements, $f_{exp}$, finite element analysis, $f_{FE}$, and Euler-Bernoulli theory, $f_{EB}$. Percentage errors in predictions of Euler-Bernoulli theory, relative to finite element results, are indicated.